# Achieving coherent perfect absorption based on flat-band plasmonic Friedrich-Wintgen BIC in borophene metamaterials

**Yan-Xi Zhang,[1] Qi Lin,[1,3] Xiao-qiang Yan,[2] Ling-Ling Wang[3] and Gui-Dong Liu,[1,3,*]**

[1]*School of Physics and Optoelectronics, Xiangtan University, Xiangtan 411105, China*
[2]*School of Mathematics and Computational Science & Hunan Key Laboratory for Computation and Simulation in Science and Engineering, Xiangtan University, Hunan 411105, China*
[3]*School of Physics and Electronics, Hunan University, Changsha 410082, China*
*[*gdliu@xtu.edu.cn](mailto:gdliu@xtu.edu.cn)*

**Abstract:** Many applications involve the phenomenon of a material absorbing electromagnetic radiation. By exploiting wave interference, the efficiency of absorption can be significantly enhanced. Here, we propose Friedrich-Wintgen bound states in the continuum (F-W BICs) based on borophene metamaterials to realize coherent perfect absorption with a dual-band absorption peak in commercially important communication bands. The metamaterials consist of borophene gratings and a borophene sheet that can simultaneously support a Fabry-Perot plasmon resonance and a guided plasmon mode. The formation and dynamic modulation of the F-W BIC can be achieved by adjusting the width or carrier density of the borophene grating, while the strong coupling leads to the anti-crossover behavior of the absorption spectrum. Due to the weak angular dispersion originating from the intrinsic flat-band characteristic of the deep sub-wavelength periodic structure, the proposed plasmonic system exhibits almost no change in wavelength and absorption at large incident angles (within 70 degrees). In addition, we employ the temporal coupled-mode theory including near- and far-field coupling to obtain strong critical coupling, successfully achieve coherent perfect absorption, and can realize the absorption switch by changing the phase difference between the two coherent beams. Our findings can offer theoretical support for absorber design and all-optical tuning.

## 1. Introduction

For many different applications, such as solar thermophotovoltaic devices [1], optical modulators [2], and optical sensors [3], the ability of materials to absorb electromagnetic radiation is critical. The absorption of the system is generally positively correlated with the near-field enhancement, which can reach a maximum of 0.5 for a single-port input in the event that the critical coupling requirement is met [4,5], i.e., the balance between the radiative and non-radiative loss rates is achieved by optimizing the structural parameters.

Under the critical coupling condition, one way to achieve perfect absorption is to add a certain thickness of the traditional precious metal plate as a Salisbury screen to block the transmission of the system [6,7]. It suffers from some drawbacks, such as the absorption and the operating frequency of the device cannot be dynamically tuned unless the device is remanufactured.

The other method is coherent perfect absorption (CPA), which is the interferometric all-optical control of absorption under the illumination of two counter-propagating input beams by controlling their intensities and relative phases [8,9]. The ability to flexibly control the absorption of light makes CPA useful for a very wide range of applications, from all-optical data processing to enhanced photocurrent generation. Studies have shown that by using materials with tunable conductivity, such as graphene or black phosphorus, as the absorbing medium based on CPA, a deep subwavelength plasmonic absorber with tunable operating frequency and absorption rate can be achieved [10,11]. Unfortunately, the plasmonic frequencies of these 2D materials are limited to the terahertz or infrared regions due to their low carrier densities, making it difficult to support plasmon resonances in commercially important communication bands. Borophene, an emerging 2D material, has unique optical properties and an electronic structure that gives it excellent potential for optical absorption [12]. Due to the higher carrier density ($\sim 10^{19}$ m$^{-2}$), borophene can support surface plasmon polaritons (SPPs) in the near-infrared region [13,14]. More importantly, the plasmonic frequency can be tuned to the communication band by designing the carrier density, which opens up the possibility of its application in the communication band.

Controlling and matching the radiative loss channel to the inherent loss of the system is essential to achieve CPA. Recently, bound states in the continuum (BICs) have gained considerable attention due to their inability to couple with the far-field, resulting in infinite radiative $Q$-factors and intensified electromagnetic near fields [15,16]. In practice, small parameter adjustments (geometric or excitation conditions) may disturb the BIC, resulting in the emergence of quasi-BICs (q-BICs) that are reachable from the far field [17,18]. Meanwhile, the radiative $Q$-factors become finite and proportional to the magnitude of the disturbance. The introduction of loss materials into this system can be very convenient to achieve critical coupling [19]. For example, the research groups of Shuyuan Xiao and Feng Wu recently presented a general method to successfully for tailoring the absorption bandwidth of graphene via critical coupling in the near-infrared region [20]. Hongjian Li and Zhimin Liu et al. investigated optical BIC in a grating-graphene-Bragg mirror structure and achieved a dual-band perfect absorber [21]. To control light in an efficient and coherent way, we proposed gap-perturbed dimerized gratings based on bulk WS2 for flexible control of self-hybridization of a quasi-BIC and excitons, and achieved polaritonic coherent perfect absorption under strong critical coupling condition [22]. However, in these studies, most of the electromagnetic energy is loosely localized in the all-dielectric resonators, i.e., its mode volume is large, which is not conducive to device integration. Guangtao Cao et al. proposed a new strategy based on F-W BICs to realize a tunable perfect absorber with a large dynamic modulation range by employing graphene metamaterials [23]. In the vicinity of the communication band, Andreas Tittl et al. proposed a novel approach for plasmonic perfect absorbers by combining mirror-coupled resonances with the unique loss engineering capability of plasmonic quasi-BICs. Based on the above studies, it is theoretically possible to realize a borophene-supported plasmonic BIC to satisfy the strong critical coupling condition and achieve CPA, and then eventually obtain a deep subwavelength plasmonic absorber with tunable frequency and absorption strength near commercially important communication bands.

In this work, we propose Friedrich-Wintgen bound states in the continuum (F-W BICs) based on borophene metamaterials to realize coherent perfect absorption with a dual-band absorption peak in commercially important communication bands. The metamaterials consist of borophene gratings and a borophene sheet which can simultaneously support a Fabry-Perot plasmon resonance and a guided plasmon mode. It is found that the formation and dynamic modulation of the F-W BIC can be achieved by adjusting the width or carrier density of the borophene grating, while the strong coupling leads to the anti-crossover behavior of the absorption spectrum. In the case of oblique incidence, due to the weak angular dispersion originating from the intrinsic flat-band characteristic of the deep sub-wavelength periodic structure, we found that the system has wide angle (within 70 degrees) and high absorption. In addition, we employ the temporal coupled-mode theory including near- and far-field coupling to obtain strong critical coupling, and successfully achieve coherent perfect absorption. It is found that by changing the phase difference between two coherent beams, we can switch between complete absorption and complete transparency to achieve absorption switching.

## 2. Design and simulations

As shown in Fig. 1(a), we propose an absorption system consisting of borophene gratings and a borophene sheet, with middle of the two separated by a silica layer. The period is $P$, the width of the borophene grating is $w$, the thickness of the silica layer is $d$, and the carrier densities of the borophene grating and the borophene sheet are $n$ and $n_0$. The structural parameters are set as follows: $P = 168$ nm, $w = 108$ nm, $d = 10$ nm, $n = n_0 = 4.3 \times 10^{19}$ m$^{-2}$, and the relative refractive index of the silica layer is 1.6. Fig. 1(b) shows the $xz$ cross-sectional view of the system in one period, which allows the structure to be observed more clearly. The finite-difference time-domain (FDTD) method is used to investigate the electromagnetic properties of the structure. In the simulation, a plane wave polarized in the $x$-direction is incident along the $z$-axis. The simulation temperature and time are set to 300 K and 20000 fs, respectively. Periodic boundary conditions and perfectly matched layers (PMLs) are set in the $x$- and $z$-directions, respectively. The grid accuracy is set to $\Delta x = 1$ nm and $\Delta z = 0.5$ nm.

The material of choice is the anisotropic borophene, and its electrical conductivity is given by the Drude formula [24]:

$$\sigma_{jj} = \frac{iD_j}{\pi(\omega+i\tau^{-1})} , \quad D_j = \pi e^2 \frac{n}{m_j} , \tag{1}$$

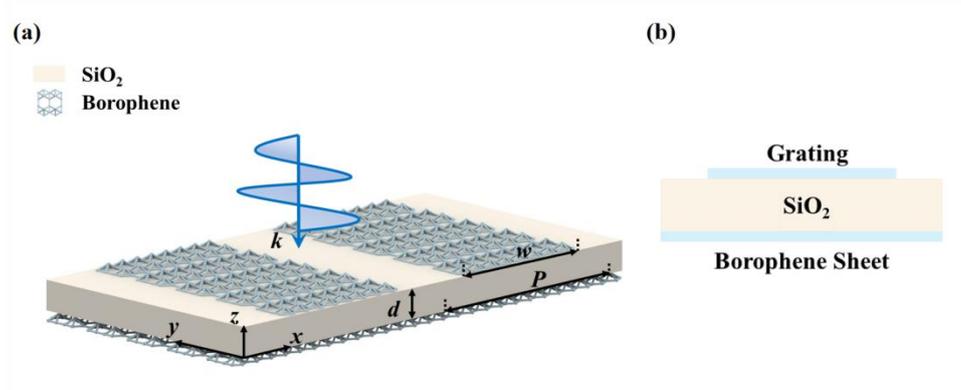

Fig. 1. (a) Schematic diagram of the absorption system (b) The *xz* cross-sectional view in one period.

where *j* represents either the *x* or *y* optical axis of the borophene crystal. The $\omega$ is the angular frequency of the incident wave, $\tau$ is the electron relaxation time set to 65 fs, *n* and *e* are the carrier density and electron charge, respectively. $D_j$ and $m_j$ stand for the Drude weight and the effective electron mass, here $m_x = 1.4\ m_0$, $m_y = 3.4\ m_0$, $m_0$ represents the standard electron rest mass.

## 3. Results and discussion

It is generally known that under suitable phase matching conditions, the destructive interference between different modes can form F-W BIC. Therefore, before analyzing the physical mechanism of F-W BIC, we first investigate the mechanism of the two modes. The absorption spectrum of the system is shown in Fig. 2(a), when the incident wave is irradiated vertically on the system, two absorption peaks with 0.5 appeared near 1399 nm and 1864 nm, where the absorption is defined by the equation $A = 1 - R - T$. Here, the $n = n_0 = 4.3 \times 10^{19}$ m$^{-2}$. Figs. 2(b) and (c) show the electric field distributions in the *z*-direction at the position of two resonance peaks. It can be seen that these are the guided mode resonance (GMR) and the Fabry-Perot resonance (FPR). The dotted line represents the position of the borophene grating and the borophene sheet. Then, we analyze the generation conditions of the two modes. With the borophene grating, which makes the incident wave can achieve the phase-matching with the guided modes in the SiO$_2$ plate, that the phenomenon is called GMR [25, 26]. The wave vector of the incident wave in the *x*-direction can be expressed by $k_x = k_0 sin\theta$, where $k_0 = \omega / c$ is the wave vector in free space, $\theta$ is the angle of incidence, $\omega$ is the angular frequency, and *c* is the speed of light in air. The GMR mode condition can be approximated as follows:

$$\text{Re}(\beta(\omega)) = |k_x + mG_x| , \tag{2}$$

here $m$ is the order of the GMR mode, we consider the case of $m = 0$. $G_x = 2\pi / P$ represents the reciprocal lattice vector caused by the borophene grating. $\beta(\omega)$ is the propagation constant of the guided mode, which can be obtained by solving the dispersion relation of the guided mode [27]:

$$\frac{n_v^2}{\sqrt{\beta(\omega)^2 - n_v^2 k_0^2}} + \frac{n_s^2}{\sqrt{\beta(\omega)^2 - n_s^2 k_0^2}} = -\frac{i\sigma_{jj}}{ck_0\varepsilon_0}, \quad (3)$$

where $n_v$ and $n_s$ are the refractive index in vacuum and the relative refractive index of the silica, respectively.

Since the impedance at the two ends is mismatched in the regions with and without borophene, the borophene surface plasmon polaritons (SPPs) are reflected back and forth at the two boundaries, forming a standing wave. The standing wave is called lateral FPR, and the frequency of the FPR mode can be written as [28]:

$$\omega_{FPR} = \frac{Kc}{2n_M w}, \quad (4)$$

where $K$ is the order of the FPR mode, $c$ is the speed of light, $n_M$ is the effective modal index of the standing wave, and $w$ is the width of the borophene grating.

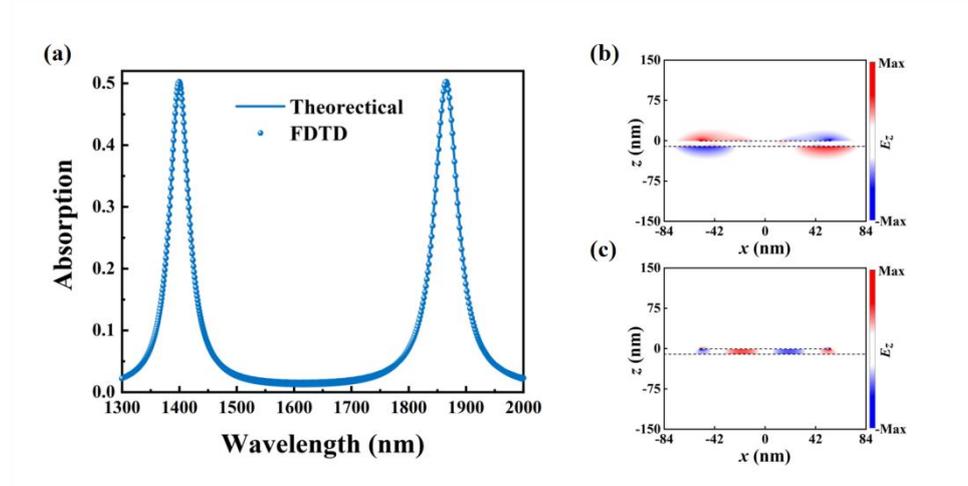

Fig. 2. (a) FDTD simulations of the absorption spectrum and results of the theoretical TCMT calculation. (b) and (c) are the electric field distributions of the two modes in the $z$-direction.

To further explore the strong coupling between the GMR mode and the FPR mode, we employ temporal coupled-mode theory (TCMT) for analysis. The dynamical equations of the two resonant modes can be expressed as [29]:

$$\frac{da_1}{dt} = (i\omega_1 - \gamma_{e1} - \gamma_{i1})a_1 + i\kappa a_2 + i\sqrt{\gamma_{e1}}s^+ + i\sqrt{\gamma_{e1}}(i\sqrt{\gamma_{e2}}e^{-i\varphi}a_2) , \qquad (5)$$

$$\frac{da_2}{dt} = (i\omega_2 - \gamma_{e2} - \gamma_{i2})a_2 + i\kappa a_1 + i\sqrt{\gamma_{e2}}s^+ + i\sqrt{\gamma_{e2}}(i\sqrt{\gamma_{e1}}e^{-i\varphi}a_1) , \qquad (6)$$

where $a_1$ and $a_2$ represent the amplitudes of the two modes, $\omega_{1(2)}$, $\gamma_{e1(2)}$ and $\gamma_{i1(2)}$ represent the eigenfrequency, external loss and internal loss of the two modes, respectively, $\kappa$ is the coupling coefficient between the two modes, $s^+$ is the incident wave, $\varphi$ is the phase difference between the two modes. By solution of Eq. (5) and Eq. (6), the amplitudes of the two modes can be expressed as:

$$a_1 = \frac{\{i(\omega-\omega_2)+\gamma_{e2}+\gamma_{i2}\}i\sqrt{\gamma_{e1}} + (i\kappa - \sqrt{\gamma_{e1}\gamma_{e2}}e^{-i\varphi})i\sqrt{\gamma_{e2}}}{\{i(\omega-\omega_1)+\gamma_{e1}+\gamma_{i1}\}\{i(\omega-\omega_2)+\gamma_{e2}+\gamma_{i2}\} - (\sqrt{\gamma_{e1}\gamma_{e2}}e^{-i\varphi} - i\kappa)^2}s^+ , \quad (7)$$

$$a_2 = \frac{\{i(\omega-\omega_1)+\gamma_{e1}+\gamma_{i1}\}i\sqrt{\gamma_{e2}} + (i\kappa - \sqrt{\gamma_{e1}\gamma_{e2}}e^{-i\varphi})i\sqrt{\gamma_{e1}}}{\{i(\omega-\omega_1)+\gamma_{e1}+\gamma_{i1}\}\{i(\omega-\omega_2)+\gamma_{e2}+\gamma_{i2}\} - (\sqrt{\gamma_{e1}\gamma_{e2}}e^{-i\varphi} - i\kappa)^2}s^+ , \quad (8)$$

and the absorption of the system is given by the formula [30]:

$$A = 2\gamma_{i1}\left|\frac{a_1}{s^+}\right|^2 + 2\gamma_{i2}\left|\frac{a_2}{s^+}\right|^2 , \qquad (9)$$

The phase difference $\varphi$ can be approximately zero due to the order of magnitude difference between the propagation velocity of the incident wave and the distance. The absorption formula derived from TCMT can be well fitted with the numerical simulation results of FDTD, as shown in Fig. 2(a). In the fitting results, the external loss and internal loss values of the two modes are consistent, i.e., both modes achieve a balance between the radiation loss rate and the non-radiation loss rate, which satisfies the critical coupling condition.

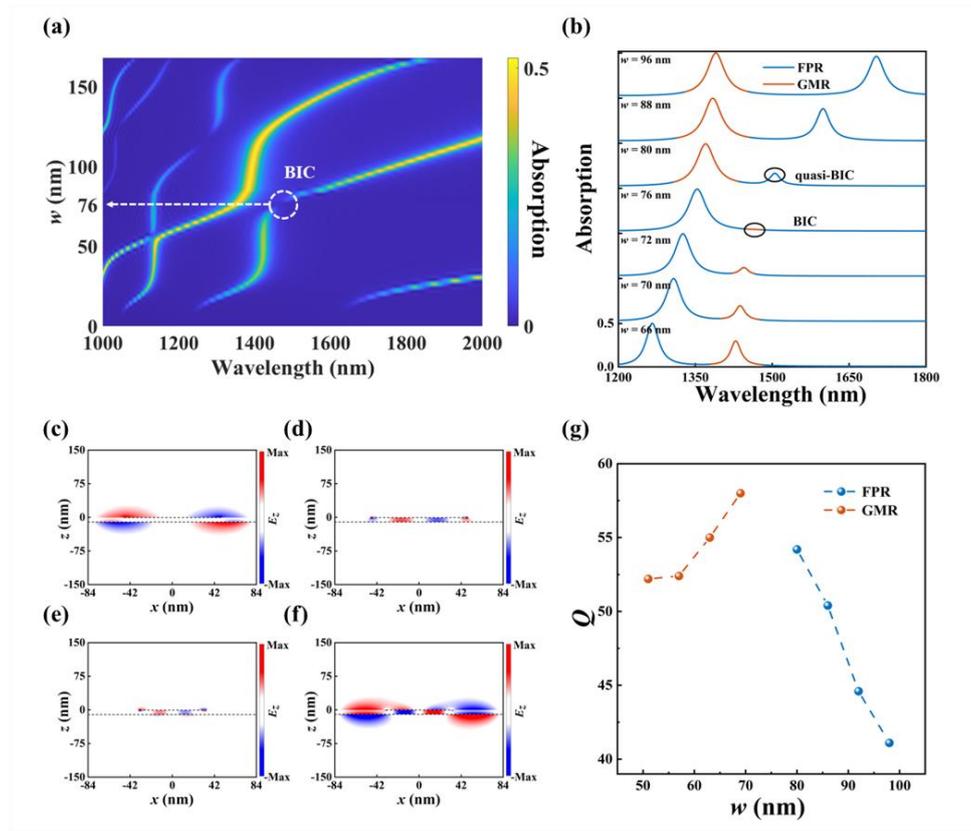

Fig. 3. The process of strong coupling of the GMR and FPR modes to form the F-W BIC (a) Absorption spectra change with the borophene grating width, here $n = n_0 = 4.3 \times 10^{19}$ m$^{-2}$. (b) Destructive interference of the GMR and FPR modes avoids crossing and linewidth vanishing. (c)-(f) The z-direction electric field distribution of the two resonance modes at $w = 96$ nm and $w = 66$ nm. (g) The variation of the Q-factor with the width of the borophene grating.

To investigate the process of strong coupling between the GMR mode and the FPR mode leading to the formation of the F-W BIC, Fig. 3(a) shows the variation trend of the absorption spectrum with the width of the borophene grating. It can be seen from the figure that the absorption spectrum has anti-crossover behavior, and the FW-BIC is formed near the frequency intersection. In order to study the interaction between the two modes, the absorption spectra under different grating widths are shown in Fig. 3(b). As the width of the borophene grating is adjusted from 96 nm to 76 nm, the resonance frequencies of the two modes gradually become equal, the quasi-BIC evolves into BIC, and the F-W BIC is formed near 1465 nm. Fig. 3(c)-(f) show the z-direction electric field distributions of the two resonance modes before and after the BIC point, it can be seen that the two modes exchange. Therefore, we can conclude that the strong coupling between the two modes makes the absorption spectrum have anti-crossover behavior, and when the resonance frequencies of the two modes are equal, the mutual interference between the modes leads to the disappearance of the line width and the formation of F-W

BIC. Next, we calculate the $Q$-factor under the coupling of the two modes using the formula $Q = f_0 / \Delta f$, where $f_0$ is the center frequency of the absorption peak and $\Delta f$ is the full width at half maximum (FWHM). The $Q$-factor diverges at $w = 76$ nm due to the absorption linewidth vanishes, which is consistent with the physical properties of the BIC point, as shown in Fig. 3(g).

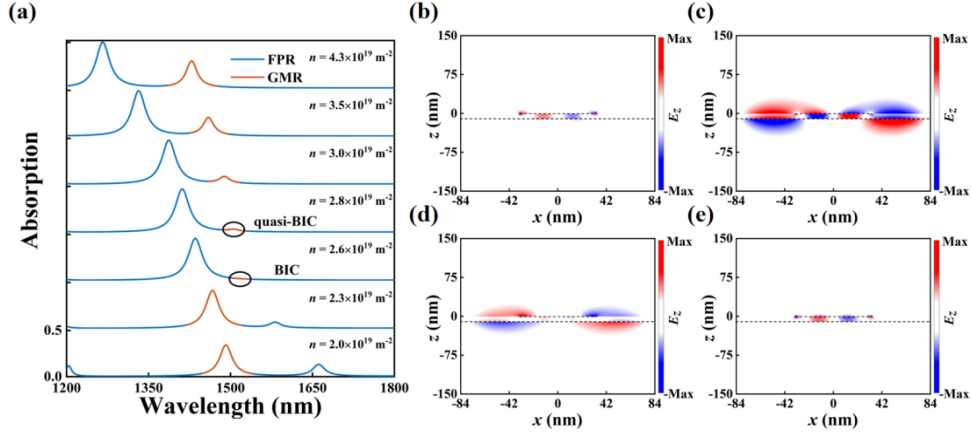

Fig. 4. (a) F-W BIC is dynamically modulated by designing the carrier density of the borophene grating ($w = 66$ nm). (b)-(e) The electric field distributions of two resonance modes in the $z$-direction at $n = 3.5 \times 10^{19}$ m$^{-2}$ and $n = 2.0 \times 10^{19}$ m$^{-2}$.

In the previous discussion, we achieved FW-BIC by adjusting the width of the borophene grating. In many studies, the dynamic modulation of FW-BIC can be achieved by changing the Fermi level or carrier density of the material by bias voltage. Therefore, we suspect that F-W BIC can be dynamically modulated by designing the carrier density of the borophene grating because of its sensitivity to carrier density. In order to verify this conjecture, simulations are performed under this condition ($w = 66$ nm). When the carrier density of the borophene grating is adjusted from $4.3 \times 10^{19}$ m$^{-2}$ to $2.0 \times 10^{19}$ m$^{-2}$, the FW-BIC is formed at $n = 2.6 \times 10^{19}$ m$^{-2}$ due to the mutual interference of the two modes. Fig. 4(b)-(e) show the electric field distributions of the two modes in the z-direction at $n = 3.5 \times 10^{19}$ m$^{-2}$ and $n = 2.0 \times 10^{19}$ m$^{-2}$, it can be observed that the two modes exchange. This confirms that the F-W BIC can be dynamically modulated by designing the carrier density of the borophene grating. It is worth mentioning that the carrier density of borophene has a wide design range.

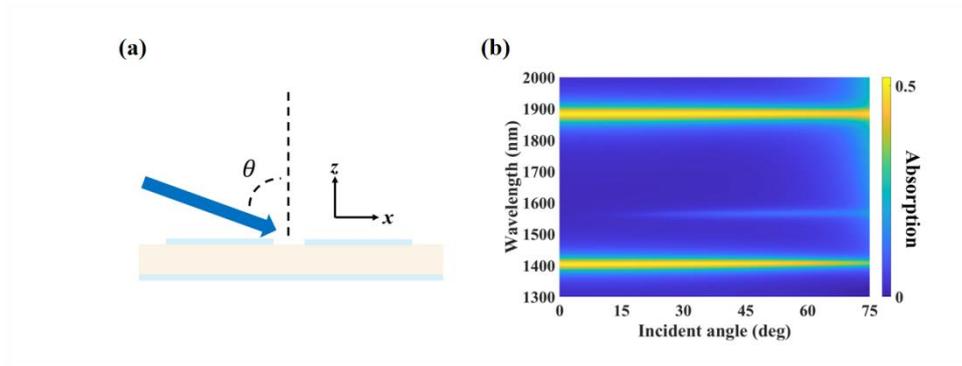

Fig.5 (a) Schematic of oblique incidence. (b) Absorption spectra change with the angle of incidence.

In addition, from a practical point of view, the influence of oblique incidence on the absorber is an important direction to evaluate the performance of the light absorber. Fig. 5(a) shows the oblique incidence model. By changing the angle of incidence of the incident wave, it is found that the absorption system can still maintain high absorption when the angle of incidence is changed to 70 degrees, as shown in Fig. 5(b). This is due to the weak angular dispersion originating from the intrinsic flat-band characteristic of the deep sub-wavelength periodic structure. The absorption system has the characteristics of insensitivity to the angle of incidence and wide-angle high absorption, which may provide the possibility of achieving wide-angle absorption.

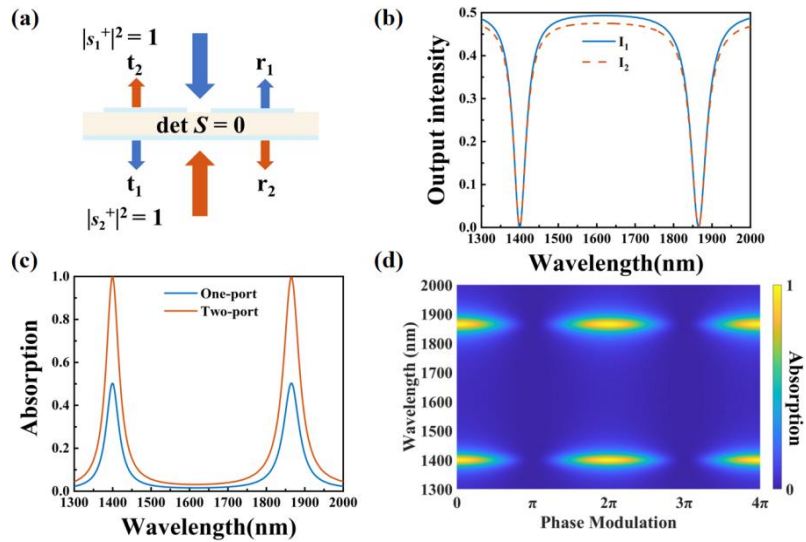

Fig. 6. (a) Schematic diagram of coherent perfect absorption. (b) The output intensities $I_1$ and $I_2$ of the system under two-port source excitation. (c) Absorption spectrum of the system under one-port and two-port excitation. (d) The dependence of the absorption spectra on the phase difference.

In order to enhance the absorption of the system, coherent perfect absorption is explored under the strong critical coupling condition, as shown in Fig. 6(a). The scattering matrix describes the relationship between the input beam and the output beam:

$$\begin{pmatrix} s_1^+ \\ s_2^+ \end{pmatrix} = \begin{pmatrix} r_1 & t_2 \\ t_1 & r_2 \end{pmatrix} \begin{pmatrix} s_1^- \\ s_2^- \end{pmatrix} \qquad (10)$$

Two coherent beams with input intensity $|s_1^+|^2 = |s_2^+|^2 = 1$ and phase difference of zero are chosen to excite the system such that the output intensity $|s_1^-|^2 + |s_2^-|^2 \approx 0$ near 1399 nm and 1864 nm, respectively, as shown in Fig. 6(b). The output intensities $I_1$ and $I_2$ are obtained from two monitors placed behind the light source. The simulation results show that the total output intensity at the resonance wavelength is almost zero, and coherent perfect absorption is achieved. The absorption reaches almost 1, as shown in Fig. 6(c). The variation trend of the system absorption as the phase difference of the two coherent beams changes from 0 to $4\pi$ is shown in Fig. 6(d). The system absorption can reach 1 when the phase differences of the two coherent beams are even numbers of $\pi$, and it is practically zero when the phase differences are odd numbers of $\pi$. The result shows that the absorption system can be dynamically tuned between complete absorption and complete transparency to achieve the switching effect. That is, another beam can be flexibly adjusted by an input beam. This method is advantageous to achieve the switching effect in specific applications.

## 4. Conclusion

In conclusion, we propose Friedrich-Wintgen bound states in the continuum (F-W BICs) based on borophene metamaterials to realize coherent perfect absorption (CPA) in commercially important communication bands. The metamaterials consist of borophene gratings and a borophene sheet, which can simultaneously support a Fabry-Perot plasmon resonance and a guided plasmon mode. The formation and dynamic modulation of the F-W BIC can be achieved by adjusting the width or carrier density of the borophene grating, while the strong coupling leads to the anti-crossover behavior of the absorption spectrum. In the case of oblique incidence, we find that the system has the characteristics of wide angle (within 70 degrees) and high absorption. In addition, based on the strong critical coupling, we realize coherent perfect absorption and absorption switching by the phase difference between two coherent beams. Our research results can provide theoretical support for the design of absorbers and all-optical tuning, and provide new ideas for the application of optical absorption.

**Funding.** This work is supported by the Scientific Research Foundation of Hunan Provincial Education Department (22B0105), Hunan Provincial Natural Science Foundation of China (2021JJ40523, 2020JJ5551), and the National Natural Science Foundation of China (62205278, 11947062, 62105276).

**Disclosures.** The authors declare no conflicts of interest.